# Variable Selection in Restricted Linear Regression Models

Y. Tuaç[1] and O. Arslan[1]

Ankara University, Faculty of Science, Department of Statistics, 06100 Ankara/Turkey

ytuac@ankara.edu.tr, oarslan@ankara.edu.tr

**Abstract**

The use of prior information in the linear regression is well known to provide more efficient estimators of regression coefficients. The methods of non-stochastic restricted regression estimation proposed by Theil and Goldberger (1961) are preferred when prior information is available. In this study, we will consider parameter estimation and the variable selection in non-stochastic restricted linear regression model, using least absolute shrinkage and selection operator (LASSO) method introduced by Tibshirani (1996). A small simulation study and real data example are provided to illustrate the performance of the proposed method for dealing with the variable selection and the parameter estimation in restricted linear regression models.

## 1. Introduction

Consider the following linear regression model

$$y = X\beta + \epsilon \qquad (1)$$

where $y$ is an $n \times 1$ vector of observations on the response variable, $X$ is a known $n \times p$ matrix of full column rank on the explanatory variables, $\beta$ is a $p \times 1$ vector of unknown coefficients and $\epsilon$ is an $n \times 1$ vector of error terms distributed as multivariate normal with mean vector zero and variance covariance matrix $\sigma^2 I$.

In general, the ordinary least squares estimation method (OLS) is used to estimate $\beta$. Under the normality assumption the maximum likelihood estimator (MLE) and the OLS estimators are the same for $\beta$. Provided that $(X^T X)^{-1}$ exists the OLS estimator is $\hat{\beta} = (X^T X)^{-1}(X^T Y)$ and it is the best linear unbiased estimator (BLUE) of $\beta$. The use of some prior information in linear regression analysis is well known to provide more efficient estimators of regression coefficients. This kind of information can be expressed in non-stochastic or stochastic restrictions. The prior information on the coefficients should be used to obtain the OLS estimator which is called restricted OLS method. In literature, there are many papers on the restricted estimation methods for the regression parameters. For instance, Groß (2003) proposed a restricted ridge regression estimator under the non-stochastic restrictions. Zhong and Yang (2007) study on a new ridge estimation method to the restricted linear model. Alheety and Kibira (2014) proposed a generalized stochastic restricted ridge regression estimator. In this paper we will consider the non-stochastic restricted regression model and combine the restricted estimation method with LASSO method to estimate the parameters and select the variables simultaneously.

## 1.1 Restricted OLS estimation for $\beta$

Suppose that the regression parameter $\beta$ has the following $m$ linear equality restrictions

$$R\beta = r, \tag{2}$$

where R is an $m \times p$ matrix of rank m and r is an $m \times 1$ vector, both consisting of known non-stochastic numbers. Now we have a restricted optimization problem define as the minimization of $(y - X\beta)^T(y - X\beta)$ under the restriction given in equation (2). This optimization problem can be solved using Lagrange method. That is, the problem is reduced to the minimization of the following function

$$Q(\beta, \mu) = (y - X\beta)^T(y - X\beta) + \mu^T(r - R\beta) \tag{3}$$

where $\mu$ is an $m \times 1$ vector of $m$ Lagrangian multipliers associated with the $m$ constraints along the $m$ rows of $R\beta = r$. First differentiate the Lagrangian with respect to $\beta$ and $\mu$ and equating these to zero gives the restricted solution for $\beta$ as follows (see Vinod and Ullah (1981)).

$$\hat{\beta}_r = \hat{\beta}_{ols} - (X^TX)^{-1}R^T[R(X^TX)^{-1}R^T]^{-1}(R\hat{\beta}_{ols} - r) \tag{4}$$

with rank $(R) = m \leq p$.

There are two challenging problems in the multiple linear regression one is the parameter estimation and the other is the selection of explanatory variables. The aim for variable selection is to construct a model that predicts well or explains the relationships in the data. There are many variable selection methods in literature. Some of them are *stepwise procedures* like: backward elimination, forward selection and stepwise regression. There are also *Criterion-based procedures* like: Mallows $C_p$ (Mallows, (1973)), The Akaike Information Criterion (AIC) (Akaike (1974)), the Bayes Information Criterion (BIC) (Schwards (1978)), and Information complexity (ICOMP) (Bozdogan (1988)). These procedures choose efficient variables after obtaining parameter estimators. Automatic variable selections like mentioned above are practically useful however they frequently show high variability and do not decrease the prediction error (e.g see Hurvich and Tsai (1990) and Breiman (1996)). The other direction in variable selection is to consider the penalized estimation procedure with some penalty function. These procedures can estimate the parameters and do the variable selection simultaneously. Some of these methods are summarized as follows. One of the popular one is the LASSO method which use the $l_1$ norm as the penalty term. LASSO method sets some of the unimportant coefficients to zero to achieve the model selection while doing the parameter estimation simultaneously. If some of the covariates are highly correlated, then the Elastic Net proposed by Zou and Yuan (2005) can be used. Smoothly Clipped Absolute Deviation (SCAD) introduced by Fan and Li (2001) and Bridge introduced by Frank and Friedman (1993) are other penalty based variable selection methods.

## 1.2 Variable Selection with LASSO

In LASSO estimation method the OLS loss function $(y - X\beta)^T(y - X\beta)$ is minimized under the restriction $\sum_{j=1}^{p} |\beta_j| \leq t$. This minimization problem can also be solved using the Langrange method. That is, the minimization of the following function with respect to $\beta$ will give the LASSO estimators for $\beta$

$$Q_N = (y - X\beta)^T(y - X\beta) + \lambda \sum_{j=1}^{p} |\beta_j| \tag{5}$$

where $\lambda > 0$ is the regularization parameter. By minimizing the above penalized least squares regression criterion, parameter estimation and variable selection can be made simultaneously. Unlike ridge regression, there is no analytic solution for the LASSO therefore numeric methods should be used.

The minimization problem in equation (5) is not differentiable, so we use the local quadratic approximation proposed by Fan and Li (2001) to avoid this problem. The penalty term can be locally approximated at $\boldsymbol{\beta}^{(0)} = (\beta_1^0, \ldots, \beta_p^0)$ by a quadratic function

$$p_{\lambda_j}(|\beta_j|) \approx p_{\lambda_j}\left(\left|\beta_j^{(0)}\right|\right) + \frac{1}{2}\frac{p'_{\lambda_j}\left(\left|\beta_j^{(0)}\right|\right)}{\left|\beta_j^{(0)}\right|}\left(\beta_j^2 - \beta_j^{(0)2}\right)$$

$$= \frac{1}{2}\frac{p'_\lambda\left(\left|\beta_j^{(0)}\right|\right)}{\left|\beta_j^{(0)}\right|}\beta_j^2 + p_\lambda\left(\left|\beta_j^{(0)}\right|\right) - \frac{1}{2}\frac{p'_\lambda\left(\left|\beta_j^{(0)}\right|\right)}{\left|\beta_j^{(0)}\right|}\beta_j^{(0)2} \qquad (6)$$

where $p'$ is the derivative of the penalty function. For the LASSO case the penalty function is defined as $p_{\lambda_j} = \sum_{j=1}^{p}|\beta_j|$. By using the local quadratic approximation at $\boldsymbol{\beta}^{(0)}$, the minimization problem near $\boldsymbol{\beta}^{(0)}$ can be reduced to a quadratic minimization problem as follows.

$$Q(\beta, \lambda) = (y - X\beta)^T(y - X\beta) + \frac{\lambda}{2}\sum_{j=1}^{p}\frac{1}{\left|\beta_j^{(0)}\right|}\beta_j^2 \qquad (7)$$

where

$$\beta_j^{(0)} \neq 0$$

Hence, at the lasso estimate $\hat{\beta}_{lasso}$, we may approximate the solution by a ridge regression form (Tibshirani (1996)),

$$\hat{\beta}_{lasso} = \left(X^TX + \lambda\Sigma_\lambda(\beta^{(0)})\right)^{-1}(X^Ty) \qquad (8)$$

where

$$\Sigma_\lambda(\beta^{(0)}) = \lambda diag\left\{\frac{1}{\left|\beta_1^{(0)}\right|}, \ldots, \frac{1}{\left|\beta_p^{(0)}\right|}\right\}.$$

In this paper we will propose restricted version of the LASSO method to estimate the parameters and select the appropriate predictors concurrently. In our proposal we combine the non-stochastic restricted regression with LASSO penalty criterion in a single objective function. That is, we will combine the minimization problems given in equation (3) and equation (7). Our main target is to estimate the unknown parameters by using prior information and do the model selection with LASSO.

The rest of paper is organized as follows. In section 2 we propose restricted LASSO regression method to achieve the parameter estimation and the variable selection simultaneously. In section 3 we provide a small simulation study with an econometric example to demonstrate the performance of proposed method. Finally we summarize the results in discussion section.

## 2. Restricted LASSO

We consider the linear regression model given in equation (1). We will estimate the parameters under the prior information given in equation (2) and select the important regressor by combining the restricted regression estimation method with LASSO. Note that in this paper we only consider the non-stochastic restrictions. Similar procedures can be carried out for the stochastic restrictions which will be our future works. To obtain the restricted LASSO regression estimators we will minimize the following objective function

$$Q_r(\beta) = (y - X\beta)^T(y - X\beta) + \mu^T(r - R\beta) + \lambda \sum_{j=1}^{p} |\beta_j|. \tag{9}$$

Equation (9) is not differentiable because of penalty function at origin; therefore, similar to the LASSO case we will use the local quadratic approximation to the penalty function. By doing so the above objective function will be reduced to the following objective function

$$Q(\beta, \mu, \lambda) = (y - X\beta)^T(y - X\beta) + \mu^T(r - R\beta) + \beta^T \Sigma_\lambda(\beta^{(0)})\beta. \tag{10}$$

Rewriting the $Q(\beta, \mu, \lambda)$ as

$$Q(\beta, \mu, \lambda) = y^T y - 2\beta^T X^T y + \beta^T X^T X \beta + \mu^T(r - R\beta) + \beta^T \Sigma_\lambda(\beta^{(0)})\beta \tag{11}$$

and taking the derivatives of (11) with respect to $\beta$ and setting to zero we get

$$\frac{\partial Q}{\partial \beta} = 0 = -X^T y + \left(X^T X + \Sigma_\lambda(\beta^{(0)})\right)\beta_r - R^T \Sigma_\lambda(\beta^{(0)}). \tag{12}$$

To solve (12) we first pre-multiply (12) with $\left(X^T X + \Sigma_\lambda(\beta^{(0)})\right)^{-1}$ and write

$$-\left(X^T X + \Sigma_\lambda(\beta^{(0)})\right)^{-1} X^T y + \beta_r - \left(X^T X + \Sigma_\lambda(\beta^{(0)})\right)^{-1} R^T \Sigma_\lambda(\beta^{(0)}) = 0. \tag{13}$$

Note that the first term in (13) is the LASSO estimator given in (8). Therefore (13) will be as follows.

$$-\hat{\beta}_{lasso} + \hat{\beta}_r - \left(X^T X + \Sigma_\lambda(\beta^{(0)})\right)^{-1} R^T \mu = 0. \tag{14}$$

Next, pre-multiplying (14) by R and using the constrain $R\beta = r$ we get,

$$\mu = \left[R\left(X^T X + \Sigma_\lambda(\beta^{(0)})\right)^{-1} R^T\right]^{-1} (r - R\hat{\beta}_{lasso}). \tag{15}$$

Finally, substituting this value of $\mu$ in (15) we obtain the restricted lasso estimate of the model parameters as follows.

$$\hat{\beta}_{rlasso} = \hat{\beta}_{lasso} - \left(X^TX + \Sigma_\lambda(\beta^{(0)})\right)^{-1} R^T \left[R\left(X^TX + \Sigma_\lambda(\beta^{(0)})\right)^{-1} R^T\right]^{-1} (R\hat{\beta}_{lasso} - r). \quad (16)$$

To sum up, if there are some prior information exists on the model parameters, by using this estimator we can achieve the parameter estimation and the model selection simultaneously in the non-stochastic restricted regression models.

## 3. Numerical studies

*3.1 Simulation*

In this section we provide a small simulation study to demonstrate the finite sample performance of the restricted LASSO. We compare the OLS without restriction, the restricted OLS, the LASSO without restriction and the restricted LASSO in terms of the variable selection and parameter estimation of the unknown regression parameters.

The simulation settings are as follows:
We generate the data from model (1) where $\boldsymbol{\beta} = (0, 1, 3, 1, 5, 0)^T$ and generate $x_i \sim N(0,1)$ The response variables are generated according to the linear regression model $y = X\beta + \epsilon$. To apply the restricted OLS and the restricted LASSO, we must have non-stochastic linear restrictions given in (2). We consider two restrictions on regression parameters with $k = 6$ as follows

$\beta_2 = \beta_4$
$\beta_3 + 2\beta_4 + \beta_5 = 1$

In order to construct a restricted structure the matrix $R$ and the vector $r$ are given as follows:
$$R = \begin{bmatrix} 0 & 1 & 0 & -1 & 0 & 0 \\ 0 & 0 & 1 & 2 & 1 & 0 \end{bmatrix}, \quad r = \begin{bmatrix} 0 \\ 10 \end{bmatrix}.$$

We generate the error terms from normal and standard *t*-distributions with 3 degrees of freedom. Standard *t*-distribution allows us to have heavy tailed error distribution. We choose the best $\lambda$ for the LASSO and the restricted LASSO methods by using cross-validation method. To investigate the effects of the outliers, we add outliers in $y$ direction and $x$ direction while $\epsilon$ is standard normal. For each case we repeated the simulation 200 times with sample sizes 50, 100 and 200.

In Tables 1-4 we summarize the simulation results, which include the percentage of correctly estimated regression models, the average of the correctly identified zero coefficients and the average of the incorrectly estimated non-zero coefficients to be zero. Also we give the mean and median of the mean squared errors (MeanMSE MedianMSE) to illustrate the performance of the parameter estimations. We will use the matrix MSE criterion. The MSE concept for an estimator can be computed as

$$MSE(\hat{\beta}) = \frac{1}{200}\sum_{j=1}^{200}(\hat{\beta}_j - \beta_j)^2$$

We repeat the process 200 times and calculate the median and the mean of this value.

There are four different cases considered, standard normal error case, *t*-distribution error case, outlier in y direction and outlier in x direction cases. We generate the outlier by using 10% of the data from $N(100,1)$.

**Table 1**

Simulation results for $N(0,1)$ error.

| n | Method | Correctly fitted | Average No. of zeros | | MeanMSE | MedianMSE |
|---|---|---|---|---|---|---|
| | | | Correct | Incorrect | | |
| 50 | OLS | 0.28 | 1.02 | 0.00 | 0.020275 | 0.015592 |
| | Res-OLS | 0.27 | 1.51 | 0.00 | 0.011959 | 0.009294 |
| | LASSO | 0.83 | 1.83 | 0.00 | 0.086370 | 0.083662 |
| | Res-LASSO | 0.93 | 1.93 | 0.00 | 0.024933 | 0.015062 |
| 100 | OLS | 0.27 | 1.02 | 0.00 | 0.008758 | 0.007574 |
| | Res-OLS | 0.27 | 1.51 | 0.00 | 0.005272 | 0.004119 |
| | LASSO | 0.96 | 1.97 | 0.00 | 0.086400 | 0.083103 |
| | Res-LASSO | 0.99 | 1.99 | 0.00 | 0.015015 | 0.011320 |
| 200 | OLS | 0.26 | 0.96 | 0.00 | 0.004465 | 0.003727 |
| | Res-OLS | 0.27 | 1.50 | 0.00 | 0.002763 | 0.002011 |
| | LASSO | 1.00 | 2.00 | 0.00 | 0.089275 | 0.085699 |
| | Res-LASSO | 1.00 | 2.00 | 0.00 | 0.012571 | 0.011052 |

**Table 2**

Simulation results for $t_3$ error.

| n | Method | Correctly fitted | Average No. of zeros | | MeanMSE | MedianMSE |
|---|---|---|---|---|---|---|
| | | | Correct | Incorrect | | |
| 50 | OLS | 0.22 | 1.00 | 0.00 | 0.050738 | 0.039949 |
| | Res-OLS | 0.25 | 1.36 | 0.00 | 0.027144 | 0.019059 |
| | LASSO | 0.57 | 1.55 | 0.00 | 0.114036 | 0.107024 |
| | Res-LASSO | 0.77 | 1.75 | 0.00 | 0.044850 | 0.021419 |
| 100 | OLS | 0.24 | 1.09 | 0.00 | 0.034774 | 0.019281 |
| | Res-OLS | 0.25 | 1.44 | 0.00 | 0.016876 | 0.009987 |
| | LASSO | 0.80 | 1.79 | 0.00 | 0.098610 | 0.086802 |
| | Res-LASSO | 0.91 | 1.90 | 0.00 | 0.032163 | 0.016629 |
| 200 | OLS | 0.30 | 1.10 | 0.00 | 0.012489 | 0.009897 |
| | Res-OLS | 0.32 | 1.46 | 0.00 | 0.007628 | 0.005853 |
| | LASSO | 0.97 | 1.98 | 0.00 | 0.086331 | 0.084640 |
| | Res-LASSO | 1.00 | 2.00 | 0.00 | 0.019915 | 0.012961 |

**Table 3**

Simulation results for $N(0,1)$ error with outlier in y direction.

| n | Method | Correctly fitted | Average No. of zeros | | MeanMSE | MedianMSE |
|---|---|---|---|---|---|---|
| | | | Correct | Incorrect | | |
| 50 | OLS | 0.02 | 0.98 | 1.88 | 399.5712 | 251.0699 |
| | Res-OLS | 0.04 | 1.01 | 1.64 | 232.9496 | 133.0213 |
| | LASSO | 0.00 | 0.15 | 0.33 | 661.1576 | 523.7337 |
| | Res-LASSO | 0.03 | 1.06 | 1.61 | 184.5306 | 99.56323 |
| 100 | OLS | 0.02 | 1.00 | 1.73 | 144.1680 | 94.38756 |
| | Res-OLS | 0.02 | 1.03 | 1.51 | 73.15030 | 43.79667 |
| | LASSO | 0.03 | 0.18 | 0.32 | 205.6319 | 158.5479 |
| | Res-LASSO | 0.03 | 1.08 | 1.52 | 59.03667 | 34.45660 |
| 200 | OLS | 0.01 | 0.99 | 1.56 | 36.37104 | 27.53325 |
| | Res-OLS | 0.05 | 1.07 | 1.15 | 20.48621 | 11.96673 |

| | LASSO | 0.00 | 0.17 | 0.41 | 54.65942 | 46.98516 |
| | Res-LASSO | 0.10 | 1.11 | 1.15 | 16.90124 | 10.66950 |

**Table 4**
Simulation results for $N(0,1)$ error with outlier in x direction.

| n | Method | Correctly fitted | Average No. of zeros | | MeanMSE | MedianMSE |
|---|---|---|---|---|---|---|
| | | | Correct | Incorrect | | |
| 50 | OLS | 0.02 | 1.99 | 1.73 | 1.428306 | 1.354666 |
| | Res-OLS | 0.80 | 1.99 | 0.20 | 0.752260 | 0.529506 |
| | LASSO | 0.00 | 0.01 | 0.25 | 3.162567 | 3.106068 |
| | Res-LASSO | 0.99 | 2.00 | 0.02 | 0.703694 | 0.434911 |
| 100 | OLS | 0.02 | 1.00 | 1.73 | 1.297586 | 1.283253 |
| | Res-OLS | 0.82 | 1.99 | 0.11 | 0.525169 | 0.437074 |
| | LASSO | 0.03 | 0.18 | 0.32 | 2.978822 | 2.954679 |
| | Res-LASSO | 0.99 | 2.00 | 0.01 | 0.450773 | 0.308602 |
| 200 | OLS | 0.00 | 2.00 | 1.99 | 1.281326 | 1.264869 |
| | Res-OLS | 0.93 | 2.00 | 0.06 | 0.235171 | 0.167161 |
| | LASSO | 0.00 | 0.00 | 0.06 | 2.908508 | 2.898449 |
| | Res-LASSO | 1.00 | 2.00 | 0.00 | 0.156157 | 0.080740 |

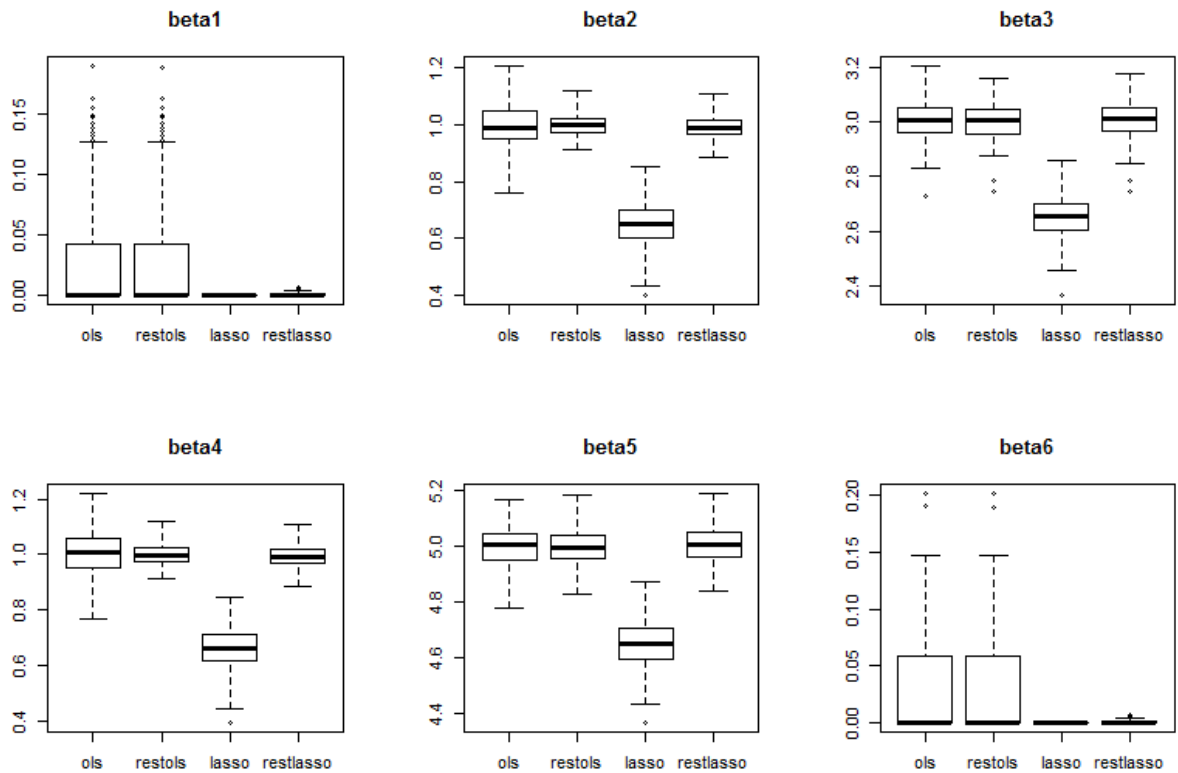

**Figure 1.** Boxplots of estimates from 200 simulated datasets in simulation study. $\epsilon$ is $N(0,1)$ with sample size 200. The thick horizontal lines show the true values of the regression coefficients.

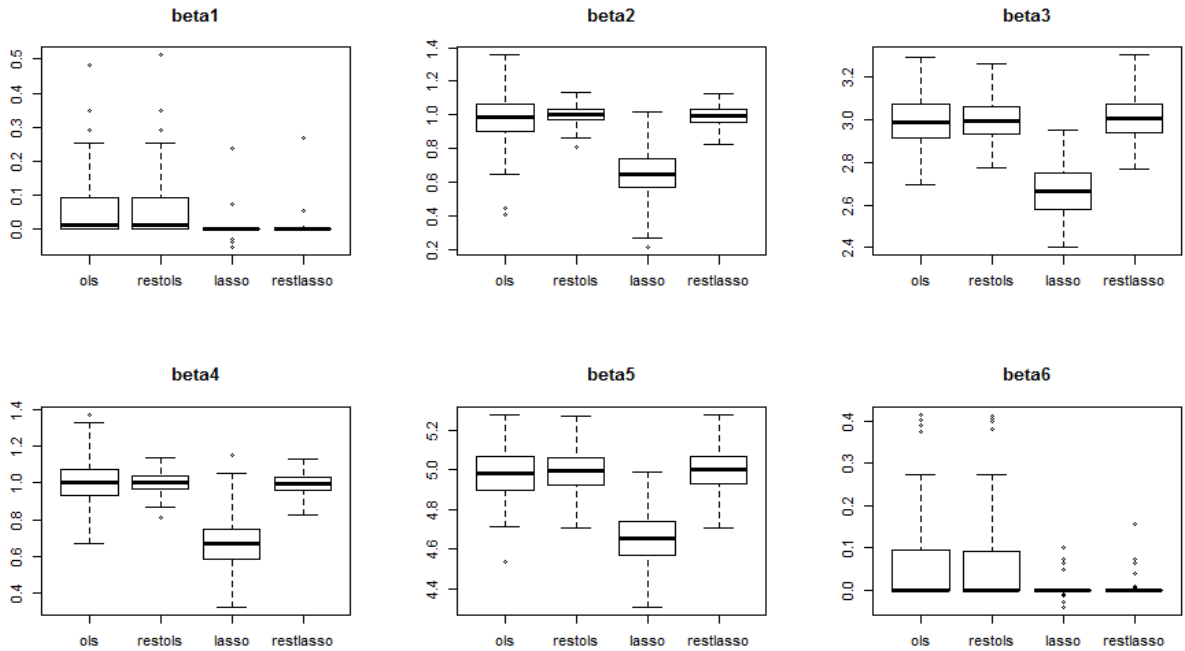

**Figure 2.** Boxplots of estimates from 200 simulated datasets in simulation study. $\epsilon$ is $t_3$ with sample size 200. The thick horizontal lines show the true values of the regression coefficients.

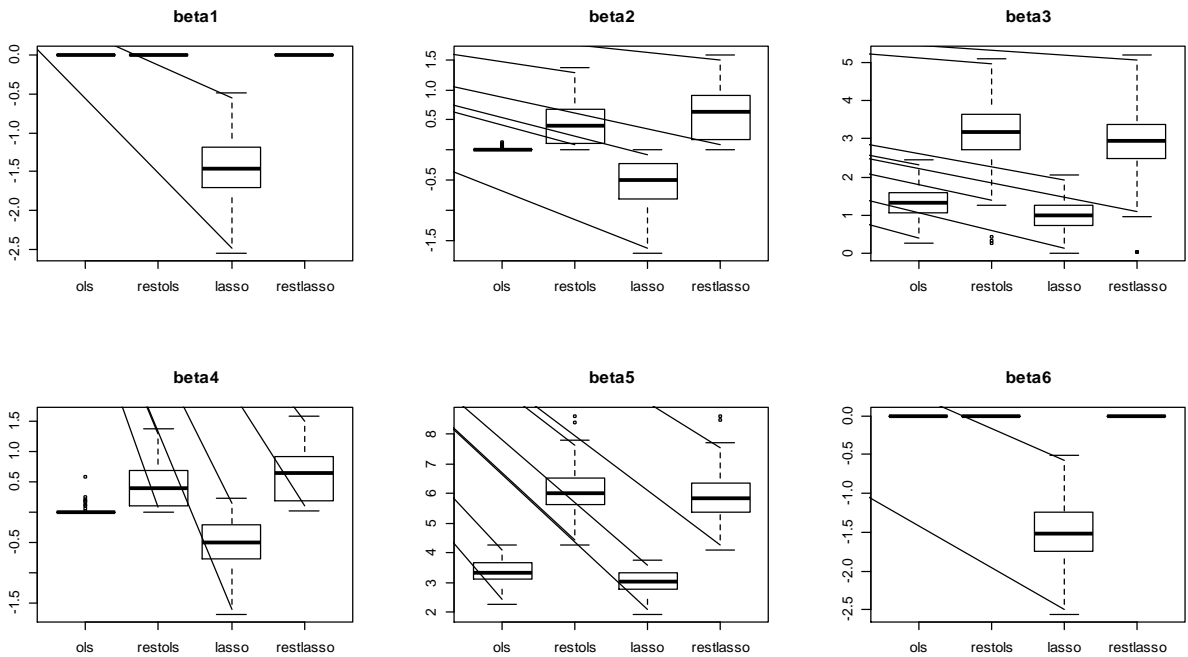

**Figure 3.** Boxplots of estimates from 200 simulated datasets in simulation study. $\epsilon$ is $N(0,1)$ with sample size 200 and outliers in x direction. The thick horizontal lines show the true values of the regression coefficients.

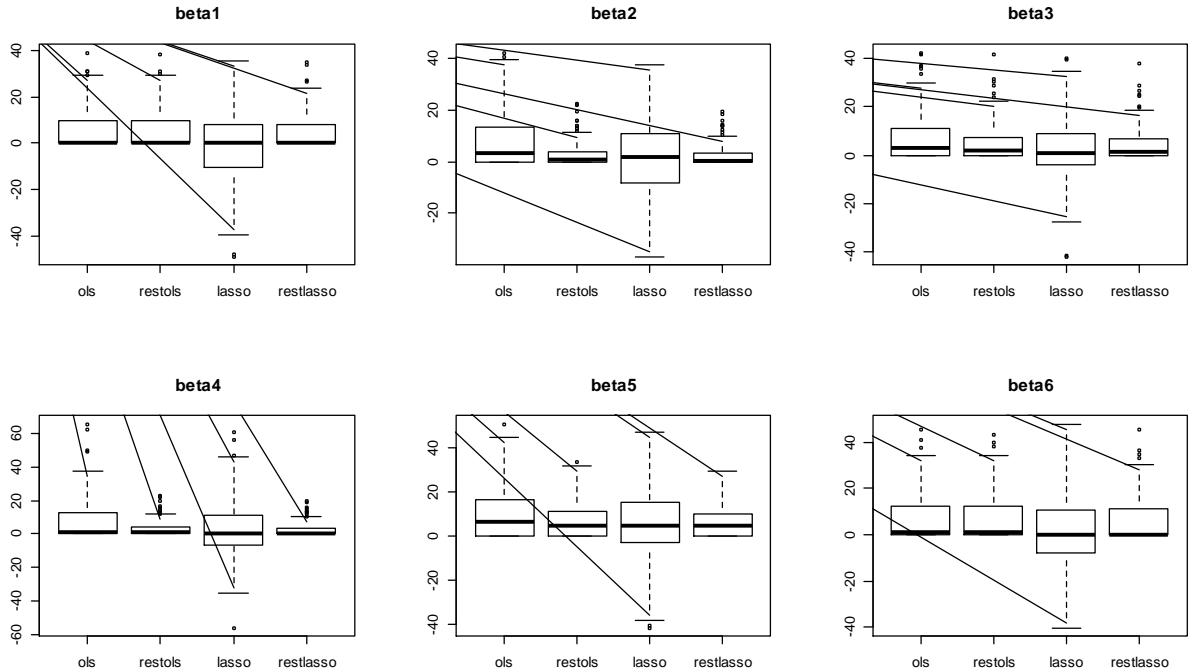

**Figure 4.** Boxplots of estimates from 200 simulated datasets in simulation study. $\epsilon$ is $N(0,1)$ with sample size 200 and outliers in y direction. The thick horizontal lines show the true values of the regression coefficients.

In first case we see that the restricted OLS has superior than the OLS method both for model selection and parameter estimation. However, the restricted LASSO method performs better than the classical LASSO in terms of model selection in case of $\epsilon$ is standard normal. The results are similar while the error distribution is standard $t$-distribution with 3 degrees of freedom. On the other hand we observe that when there are some outliers in x direction the restricted LASSO method performs better than the OLS, LASSO and restricted OLS in model selection case. The restricted LASSO method can correctly identify the number of zero coefficients and has the smallest average and median MSEs among the other methods. In this case the LASSO method cannot select the true model in any sample sizes, however the restricted LASSO method can correctly obtained the correct model in every sample sizes. The worst performance of the restricted LASSO method can be observed for the case when the error distribution is standard normal and there are outliers in y direction.

Figures 1-4 are the boxplots of the regression estimates from 200 simulated datasets for the sample size 200 in all four cases. In Figure 1 we observe that the variabilities of the estimates obtained from the LASSO and restricted LASSO are not different under the normal error case. Also all the methods are truly estimate the regression parameters. Figure 2 shows quite similar results beside some biases. Figure 3 shows that if there are some outliers in x direction the LASSO fails to estimate the true value of the parameters. Surprisingly restricted LASSO is not affected by the existence of the outliers in x direction. From Figure 4 we can say that if there are some outliers in y direction we need to use robust estimation methods to get estimators that are not affected by the outliers. That is we should combine the robust and the restricted regression methods with the LASSO to achieve parameter estimation, model selection and the restriction simultaneously.

## 3.2 Real Data Example

We use the data set has been previously analyzed by Gruber (1998) Akdeniz and Erol (2003) and Zhong and Yang (2007) to show the performance of different types of restricted regression methods. Here we use the data to evaluate the performance of both the restricted regression and the variable selection methods.

The data set consist of 10 observations observed on the four explanatory variables and one response. The response variable $Y$ is the percentage spent by United States and the explanatory variables are the percentages spent by France, West Germany, Japan and Soviet Union, respectively. Table 5 gives Total National Research and Development Expenditures as a Percent of Gross National Product by Country.

Table 5. Total National Research and Development Expenditures as a Percent of Gross National Product by Country

| Year | $Y$ | $X_1$ | $X_2$ | $X_3$ | $X_4$ |
|---|---|---|---|---|---|
| 1972 | 2.3 | 1.9 | 2.2 | 1.9 | 3.7 |
| 1975 | 2.2 | 1.8 | 2.2 | 2.0 | 3.8 |
| 1979 | 2.2 | 1.8 | 2.4 | 2.1 | 3.6 |
| 1980 | 2.3 | 1.8 | 2.4 | 2.2 | 3.8 |
| 1981 | 2.4 | 2.0 | 2.5 | 2.3 | 3.8 |
| 1982 | 2.5 | 2.1 | 2.6 | 2.4 | 3.7 |
| 1983 | 2.6 | 2.1 | 2.6 | 2.6 | 3.8 |
| 1984 | 2.6 | 2.2 | 2.6 | 2.6 | 4.0 |
| 1985 | 2.7 | 2.3 | 2.8 | 2.8 | 3.7 |
| 1986 | 2.7 | 2.3 | 2.7 | 2.8 | 3.8 |

The observations are annually from the 1972-1986. Zhong and Yang (2007) used a restricted ridge regression estimator with a non-stochastic restriction $R\beta = r$ with

$$R = \begin{bmatrix} 1 & 1 & 1 & 1 \\ 0 & 1 & 3 & 1 & 2 \end{bmatrix} \quad \text{and} \quad r = \begin{bmatrix} 1.2170 \\ 1.0904 \end{bmatrix}.$$

We use the same restrictions that Zhong and Yang (2007) used. The data are investigating by using the formula;

$(y_{j+1} - y_j)/(X_{i,j+1} - X_{i,j})$ therefore computed values are averaged over $j$ for each variable $X_i$ and we produce prior information as follows: $0.7 \quad 0.5 \quad 0.43 \quad -0.10$ for $X_1, X_2, X_3$ and $X_4$ respectively. Therefore these values are used as prior information $\beta_0 = [0.6 \quad 0.7 \quad 0 \quad 0.6 \quad -0.5]'$

After obtaining this prior information for the data we calculate the restricted OLS and our proposed method restricted LASSO. To compare the performance of the proposed estimator we calculate the estimated mse values.

Table 6. Selected variables for the data set.

| Method | Selected variables |
|---|---|
| OLS | (1,3,4) |
| Res-OLS | (1,2,3,4) |
| LASSO | (1,3) |
| Res-LASSO | (1,3) |

Table 6 shows the variable selection results using the whole data set obtained from the OLS, Restricted OLS, LASSO and Restricted LASSO methods. All the methods chose the variables $X_1$ and $X_3$. The LASSO and the restricted LASSO methods select the same variables therefore; we can say that the Restricted LASSO has the same performance according to selection of the variables with LASSO method.

We summarize the estimated mse values; 0.077081, 0.0402992, 0.214552 and 0.035209 for OLS, Res-OLS, LASSO and restricted-LASSO respectively. Since the restricted LASSO estimator has the smallest mse value among the others it has superior over the OLS, Res-OLS and LASSO estimators in terms of estimating the parameters and selecting the variables.

## 4. Conclusions and Discussion

In this paper, we have proposed a variable selection procedure for the restricted regression models with non-stochastic restriction. Our variable selection proposal based on the well-known LASSO method. The Restricted-LASSO method combines the restricted regression and LASSO estimation methods. By combining these two methods we have achieve both model selections and estimated the unknown parameters with less bias then the non-restricted LASSO. We have use the local quadratic approximation procedure proposed by Fan and Li (2001) to solve the minimization problem to obtain the estimator.

We have provided a simulation study and a real data example to illustrate the performance of the restricted-LASSO method in terms of parameter estimation and variable selection. Simulation study shows that the Restricted-LASSO method behaves well in terms of variable selection and parameter estimation when there is a non-stochastic restriction exists on the model parameters. Also we have shown that if there are some outliers exist on the x direction, both the variable selection and parameter estimation case the restricted LASSO method can handle this problem better than the other methods. However, if the outliers exist on y direction all the methods have failed therefore, we need to combine some robust method to overcome this issue. By using the robust methods we can make variable selection and robust parameter estimation simultaneously. This study will be our further consideration.

**References**

Akaike, H. (1973). Information theory and extension of the maximum likelihood principle. In B.N Petrov and F. Csáki (Eds.), Second international symposium on information theory, Académiai Kiadó, Budapest, 267-281.

Akdeniz, F. and Erol, H., Mean squared error matrix comparisons of some biased estimators in linear regression, Commun. Stat. Theory Methods 32 (2003), pp. 2389–2413.

Alheety, M. I. and Kibria, B. M. G., (2014) A Generalized Stochastic Restricted Ridge Regression Estimator, Communications in Statistics - Theory and Methods, 43:20

Bozdogan, H. (1988). ICOMP: a new model-selection criterion. In Classification and Related Methods of Data Analysis, H. H. Bock (Ed.), Elsevier Science Publishers,Amsterdam; 599-608.

Breiman, L. (1996) Heuristics of instability and stabilization in model selection. Ann. Statist., 24, 2350–2383.

Fan, J and Li, R. (2001). Variable selection via Nonconcave Penalized Likelihood and Its Oracle Properties. Journal of the American Statistical Association, Vol.96, No. 456, pp. 1348-1360

Frank, I.E. ve Friedman, J.H., (1993) A statistical view of some chemometrics regression tools, Technometrics 35, pp. 301-320.

Groß, J., Restricted ridge estimation, Statistics & Probability Letters 65 (2003) 57-64.

Gruber, M.H.J. Improving Efficiency by Shrinkage, Marcel Dekker, Inc, NewYork, 1998.

Hurvich, C. M. and Tsai, C. L. (1990) The impact of model selection on inference in linear regression.

Schwarz, G. (1978). Estimating the dimention of a Model. Annual Statist. 6, 461-4.

Tibshirani, R. (1996). Regression Shrinkage and Selection via The Lasso. Journal of the Royal Statistical Society Series B, 58:267-288.

Theil, H. and Goldberger, A. S. (1961). International Econometric Review, Vol. 2, No. 1, pp. 65-78.

Vinod, D. H. and Ullah, A. (1981). Recent Advances in Regression Methods. Marcel Dekker Inc. New York.

Zhong, Z. and Yang, H., Ridge estimation to the restricted linear model, Commun. Stat. Theory Methods 36 (2007), pp. 2099–2115.Amer. Statist., 44, 214–217.

Zou, H. and Yuan, T. (2005) Regularization and variable selection via the elastic net. J.R. Statist. Soc. B. 67, part 2, pp. 301-320.